\input epsf.sty

\documentclass[twocolumn]{revtex4}

\usepackage{amsmath}
\usepackage[english]{babel}
\usepackage[T1]{fontenc}
\usepackage{amsfonts}
\usepackage{amssymb}
\usepackage[dvips]{graphicx}

\begin{document}

\title{\textbf{New method to study stochastic growth equations: a cellular automata perspective}}

\author{ T.G. Mattos$^{1,2}$, J.G. Moreira$^1$, and A.P.F. Atman$^3$}

\affiliation{\vspace{0.5cm}$^1$Departamento de Física, Instituto de Ciências Exatas, Universidade Federal de Minas Gerais,\\C.P. 702, 30161-123 Belo Horizonte MG, Brazil\\ \\$^2$Instituto de Física, Universidade Federal Fluminense,Avenida Litorânea s/n, 24210-340 Niterói RJ, Brazil\\ \\$^3$Centro Federal de Educação Tecnológica (CEFET-MG), 30510-000 Belo Horizonte MG, Brazil}

\date{\today}

\begin{abstract}

\noindent
We introduce a new method based on cellular automata dynamics to study stochastic growth equations. The method defines an interface growth process which depends on height differences between neighbors. The growth rule assigns a probability $p_{i}(t)=\rho$~exp$[\kappa~\Gamma_{i}(t)]$ for a site $i$ to receive one particle at a time $t$ and all the sites are updated simultaneously. Here $\rho$ and $\kappa$ are two parameters and $\Gamma_{i}(t)$ is a function which depends on height of the site $i$ and its neighbors. Its functional form is specified through discretization of the deterministic part of the growth equation associated to a given deposition process. In particular, we apply this method to study two linear equations - the Edwards-Wilkinson (EW) equation and the Mullins-Herring (MH) equation - and a non-linear one - the Kardar-Parisi-Zhang (KPZ) equation. Through simulations and statistical analysis of the height distributions of the profiles, we recover the values for roughening exponents, which confirm that the processes generated by the method are indeed in the universality classes of the original growth equations. In addition, a crossover from Random Deposition to the associated correlated regime is observed when the parameter $\kappa$ is varied.

\vskip 2pc
\noindent{PACS number(s): 89.75.Da, 02.50.-r, 68.35.Ct, 05.10.-a  }

\end{abstract}

\maketitle

\section{Introduction}

Discrete computational growth models have been largely investigated along the last decades, due to the great interest in describing various features of interface growth phenomena, observed in a wide range of physical processes~\cite{fv livro, barab, meakin}. As examples of kinetic roughening models we can mention the Eden model~\cite{eden}, the ballistic deposition model (BD)~\cite{db} and some solid-on-solid growth models in which correlation mechanisms are present, such as surface relaxation~\cite{family}, height difference restriction~\cite{kk}, curvature restriction~\cite{kim-das sarma} and surface diffusion~\cite{wolf villain,das sarma tamborenea}. From an experimental point of view, growing interfaces can be generated, for example, by Molecular Beam Epitaxy (MBE) and vapor deposition over cold substrates (see Refs.~\cite{meakin,barab} and references therein).

The methodology used to study interface growth phenomena have also been applied to investigate Cellular Automata (CA)~\cite{wolfram}, an immense class of computational models which describe many phenomena in a wide sort of scientific subjects. By means of an accumulation method~\cite{sales}, usually one can map the evolution of a CA into a growing profile and then apply the tools used to analyze such systems. This method has been used recently to study deterministic~\cite{bjp1} and probabilistic CA~\cite{allbens}.

In the analysis of interface growth one is generally concerned about the temporal behavior of the interface roughness $\omega(L,t)$, which is defined at time $t$ as

\begin{equation}\label{definicao rugosidade}
\omega^2(L,t) = \frac{1}{L}\sum_{i=1}^{L}\left[ h_{i}(t)-\overline{h}(L,t)\right]^2 ~,
\end{equation}

\noindent
where $L$ is the size of the substrate and $\overline{h}(L,t)$ is the mean height of the generated profile.

For growing systems it is known that the roughness grows initially with time $t$ as a power law, defining the growth exponent, $\beta$. After a saturation time $t_{\rm x}$, the interface reaches a stationary regime and the roughness saturates. Both the saturation roughness, $\omega_{sat}$, and saturation time, $t_{\rm x}$, depend on the system size, $L$, as a power law, defining the roughness exponent, $\alpha$, and the dynamic exponent, $z$, respectively. The roughness of the surface follows the Family-Vicsek scaling law~\cite{fv}

\begin{equation} \label{scaling law}
\omega(L,t)\sim L^{\alpha}f\left( \frac{t}{L^z}\right),
\end{equation}

\noindent
where the scaling function $f(u)$ behaves as $f(u)\sim u^{\beta}$ for $u\ll 1$, and $f(u) = $ constant for $u\gg 1$. It follows that $\beta = z/\alpha$.

A set of values for the roughening exponents, in a given dimension, specifies a Universality Class (UC). Thus, if two or more processes have the same exponents values, one can say that they belong to the same UC, which means that their underlying dynamics have the same symmetries and conservation laws.

Based on these symmetries, one can write a stochastic differential equation and associate it to the given UC. This continuum approach often provides an analytical treatment for growing interfaces and exact values to the roughening exponents can, sometimes, be obtained.

In this paper we consider two linear equations and one non-linear one. They are, respectively, the Edwards-Wilkinson (EW) equation~\cite{ew},

\begin{equation}\label{equacao ew}
\frac{\partial h(\textbf{x},t)}{\partial t}~=~\nu\nabla^{2}h(\textbf{x},t)+\eta(\textbf{x},t)~,
\end{equation}

\noindent
the Mullins-Herring (MH) equation~\cite{herring, mullins},

\begin{equation}\label{equacao mh}
\frac{\partial h(\textbf{x},t)}{\partial t}~=~-K\nabla^4h(\textbf{x},t)+\eta(\textbf{x},t)~,
\end{equation}

\noindent
and the Kardar-Parisi-Zhang (KPZ) equation~\cite{kpz},

\begin{equation}\label{equacao kpz}
\frac{\partial h(\textbf{x},t)}{\partial t}=\nu\nabla^{2}h(\textbf{x},t)+ \dfrac{\lambda}{2}\left( \nabla h \right)^{2} + \eta(\textbf{x},t)~.
\end{equation}

\noindent
Here, $h(\textbf{x},t)$ is the local height of the profile, assumed to be continuous (eventual hangovers are to be ignored), and $\eta(\textbf{x},t)$ is a gaussian noise, with zero mean and correlation given by

\begin{equation} \label{noise}
\left\langle \eta\left( \textbf{x},t \right) \eta\left( \textbf{x}^{\prime},t^{\prime} \right) \right\rangle = 2D\delta^{d}\left( \textbf{x}-\textbf{x}^{\prime} \right) \delta\left( t-t^{\prime} \right).
\end{equation}

The linear equations can be solved exactly and the values of the roughening exponents can be determined. For the UC associated to the EW equation \eqref{equacao ew}, the exponents for a $d$-dimensional substrate are~\cite{ew, nattermann}

\begin{equation} \label{ew exponents}
\alpha=\frac{2-d}{2}~~,~~~\beta=\frac{2-d}{4}~~~\textrm{and}~~~z=2~.
\end{equation}

This equation can be associated to the Random Deposition with Surface Relaxation model (RDSR)~\cite{family}, where the particles, after being deposited in a random position on the lattice, are allowed to relax to the local minimum, considering the nearest neighborhood of the chosen site, and sticks irreversibly to the aggregate.

The solution for the MH equation \eqref{equacao mh} yields~\cite{wolf villain,das sarma tamborenea}

\begin{equation} \label{mh exponents}
\alpha=\frac{4-d}{2}~~,~~~\beta=\frac{4-d}{8}~~~\textrm{and}~~~z=4~.
\end{equation}

\noindent
This equation is associated to a deposition model known as Growth with Surface Diffusion (GSD), where the newly arrived particle seeks the position in the neighborhood where the bonding is maximum~\cite{wolf villain,das sarma tamborenea}.

On the other hand, non-linear equations have no general solutions for the roughening exponents. Nevertheless, for the KPZ equation \eqref{equacao kpz} with $d=1$, renormalization group theory provides~\cite{kpz}

\begin{equation} \label{kpz exponents}
\alpha=\frac{1}{2}~~,~~~\beta=\frac{1}{3}~~~\textrm{and}~~~z=\frac{2}{3}~.
\end{equation}

\noindent
This equation can be associated to two discrete models: the Ballistic Deposition (BD)~\cite{db} and the Restricted Solid-on-Solid growth model (RSOS)~\cite{kk}. In the BD, the falling particle sticks to a vertical position such that

\begin{equation}
h_i(t+1)~=~\textrm{max}~[ h_{i-1}(t), h_i(t)+1, h_{i+1}(t) ]~.
\end{equation}

\indent
In the RSOS model, at each time step, a random position is chosen in the lattice and its height is increased by one unit, provided the restriction $\left| \Delta h \right| \leq m$ is obeyed, where $\Delta h\equiv h_i-h_{i\pm 1}$.

Crossovers between distinct UCs is a topic of great interest as well. Much attention have been given to competitive growth models, where two different types of particles are deposited, one with probability $P$ and the other one with probability $(1-P)$, allowing several combinations to crossover between different growth processes~\cite{fabio2,braunstein}. In particular, models where some kind of correlated growth occurs with probability $P$ and Random Deposition (RD) with probability $(1-P)$ have been investigated recently, by means of simulations and scaling arguments~\cite{novotny, fabio1}. A crossover from random to correlated growth in the RSOS model have also been studied recently by Aarão Reis~\cite{fabio1} and a crossover from KPZ to EW class has been obtained by da Silva and Moreira in a growth model with restricted surface relaxation~\cite{tales}.

Our purpose is to develop a method~\cite{bjp2} to study those equations, based on CA dynamics. By using a simple discretization scheme, we obtain numerical solutions for the roughening exponents without actually having to solve it, which means that the method can be applied to equations whose solutions are not even known. In section \ref{method} we outline the main features of the method such as definition of the parameters and their range of interest, discretization schemes and a brief description of the algorithm. In section \ref{exponents} we show numerical results obtained for the roughening exponents corresponding to each UC considered. In section \ref{crossover kappa} we present a description of the crossover from Random Deposition (RD) to correlated growth obtained by variation of $\kappa$. Finally, in section \ref{conclusions}, we draw some conclusions and perspectives for further works.


\section{The method} \label{method}

Consider a one-dimensional lattice of size $L$, initially flat and with periodic boundary conditions. In each time step, all the sites are simultaneously visited and site $i$ receives a particle with probability $p_i(t)$, given by

\begin{equation} \label{definicao do modelo}
p_i(t)=\rho~e^{\kappa\Gamma_i(t)}~.
\end{equation}

\noindent
Here $0<\rho<1$ and $\kappa>0$ are two parameters, fixed throughout the evolution of the interface. The former is related to the growth speed and the later can be associated to an inverse of temperature if one considers an analogy with vapor deposition: the synchronous update scheme can be thought as attempts for deposition of vapor particles on a cold substrate, allowing the development of growing structures. $\Gamma_i(t)$ is the kernel, a function which depends on heights of site $i$ and its neighborhood, at time $t$. Its explicit form will be given by the discretization of the deterministic part of the growth equation we are intended to study.

In the case of the EW equation \eqref{equacao ew}, the kernel is given by the discretization of the Laplacian $\nabla^2h$,

\begin{equation} \label{kernel ew}
\Gamma_i(t) = h_{i+1}(t)+h_{i-1}(t)-2h_i(t)~.
\end{equation}

For the MH equation \eqref{equacao mh}, the kernel follows from the discretization of the negative of the fourth spatial derivative, $-\nabla^4h$,

\begin{eqnarray} \label{kernel dif}
\Gamma_i(t) & = & -\left[ 6h_i(t) + h_{i+2}(t) + h_{i-2}(t)\right] +\nonumber \\
 & & +~4\left[ h_{i+1}(t) + h_{i-1}(t) \right]~.
\end{eqnarray}
\\
Finally, the discretization of the square of the gradient, $(\nabla h)^2$, and the Laplacian yields the kernel for the KPZ equation \eqref{equacao kpz}, which is

\begin{eqnarray} \label{kernel kpz}
\Gamma_i(t) & = & \frac{1}{\varepsilon}\left[ h_{i+1}(t)-h_{i-1}(t)\right]^2 +\nonumber \\
 & & +\left[ h_{i+1}(t)+h_{i-1}(t)-2h_i(t)\right]~,
\end{eqnarray}

\noindent
\\
where $\varepsilon >0$ is the parameter which controls the non-linearity strength: large $\varepsilon$ implies a small contribution of the non-linear term, and conversely.

So far we have explicitly considered only the deterministic part of the growth differential equations \eqref{equacao ew} to \eqref{equacao kpz}. Nevertheless, the stochastic nature of such equations contained in the gaussian noise $\eta(\textbf{x},t)$ is simulated in our method by the probabilistic character of the growth process. We have done a rigorous study of the symmetry and decay properties of the height distributions, which corresponded to those of a gaussian distribution.

Note that, in the way we have defined the method, one can eventually obtain $p_i(t) > 1$. For this situation, we impose the condition

\begin{equation}\label{condicao}
p_i(t)\geq 1 ~ \Longrightarrow ~ p_i(t)=1 ~ \Longrightarrow ~ h_i(t+1)=h_i(t)+1~.
\end{equation}

Hence, given a pair of values $(\rho,\kappa)$, there is a maximum kernel value, $\Gamma_{max}(\rho,\kappa)$, for which

\begin{equation}
\Gamma_i(t) \geq \Gamma_{max} ~\Longrightarrow ~ p_i(t)=1~.
\end{equation}

Making $p_i(t)=1$ in equation \eqref{definicao do modelo} and having in mind the fact that $\Gamma_i(t)$ is an integer by definition, we find

\begin{equation} \label{gamma max}
\Gamma_{max}(\rho,\kappa)=\textrm{int} \left( - ~ \frac{1}{\kappa} ~ \ln\rho \right) ~.
\end{equation}

Basically, for each time step $t$ the algorithm does the following: \textbf{(i)} calculate $\Gamma_i(t)$; \textbf{(ii)} ask whether $\Gamma_i(t) \geq \Gamma_{max}$; \textbf{(iii)} if so, a particle should be deposited in site $i$; if not, a random number $r$ is taken in the range $[0,1)$ and $p_i(t)$ is calculated: if $r < p_i(t)$, a particle should be deposited on site $i$; \textbf{(iv)} repeat steps (i) to (iii) for $i=1,\ldots,L$; \textbf{(v)} those sites that should receive a particle, have their heights simultaneously increased by one unit.

For consistency between the algorithm and the definition of the method, equation \eqref{definicao do modelo}, we must have $\Gamma_{max}\geq 1$, because $\Gamma_{max} = 0$, according to the algorithm, would make the interface grow flat since the beginning of the process and no scaling features would be observed. This condition restricts the range of values of $\kappa$ that are to be considered,

\begin{equation}\label{condicao kappa}
\Gamma_{max}\geq 1 ~\Longrightarrow ~\kappa \leq \ln\left( \frac{1}{\rho} \right)~.
\end{equation}

We fixed $\rho = \frac{1}{2}$ throughout our simulations so that, in a flat surface, each site has probability $p=\frac{1}{2}$ to receive a particle. Further analysis of the displacement rate of the mean height showed that the parameter $\rho$ is related to the growth speed of the interface.


\section{Roughening exponents} \label{exponents}

In this section we show the results obtained for the roughening exponents, for each one of the three UCs. In a lattice of size $L=10^4$, we initially fixed $\kappa=10^{-1}$ and averaged the results over $50$ independent samples.

\begin{figure}[h]
\begin{center}

\includegraphics[width=6.67cm , height=6cm]{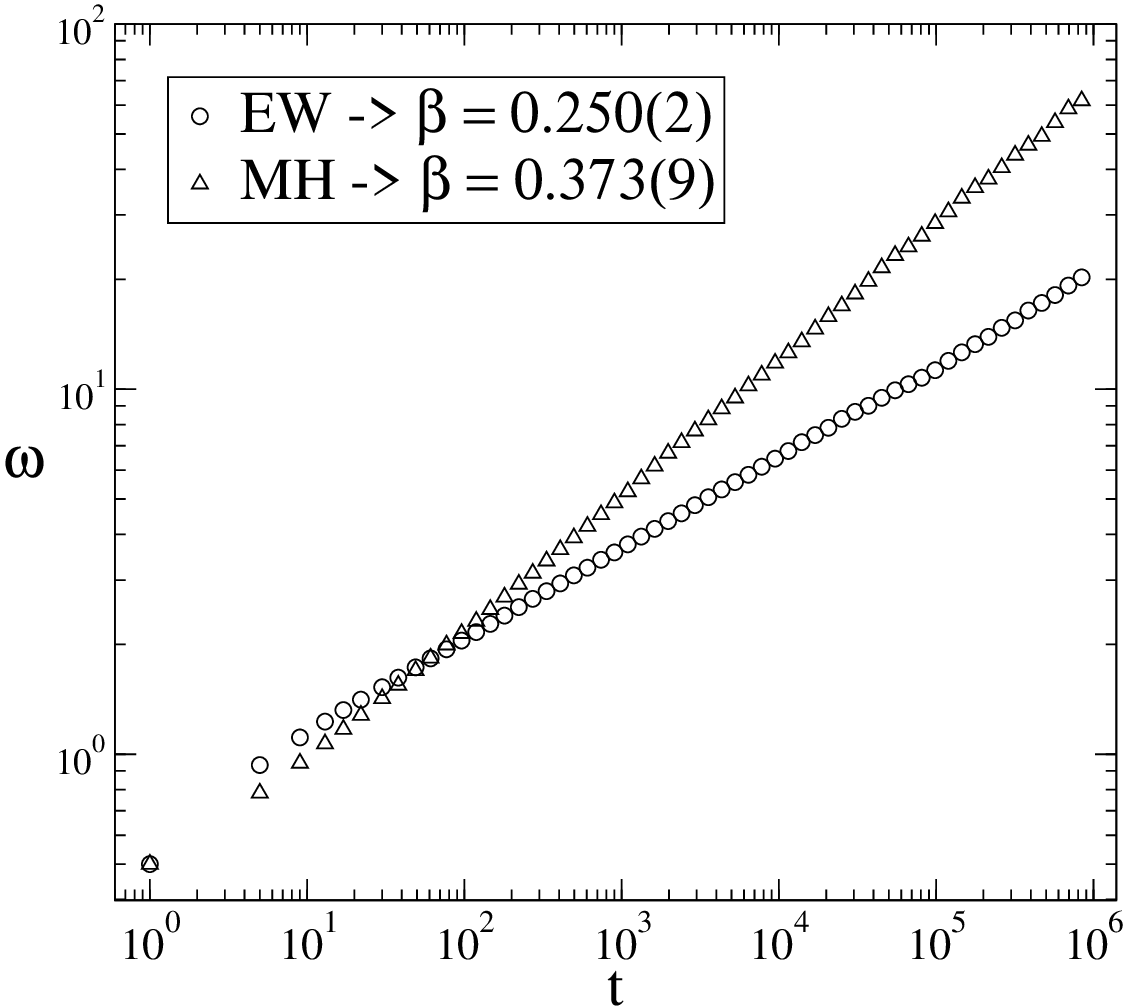}

\caption{\textit{\small Log-log plots of the roughness as function of time, for $\rho=1/2$, $\kappa=10^{-1}$, $L=10^4$ and averaged over 50 independent samples. The circles represent the data for the EW class ($\beta=1/4$ expected), while the triangles are the data for the MH class ($\beta=3/8$ expected). Least squares method was used in order to determine the indicated numerical values for $\beta$.}}

\label{rough ew-gsd}

\end{center}
\end{figure}

The results obtained for the roughness of the interfaces generated by the application of the method to the EW and MH equations are shown in figure \ref{rough ew-gsd}. Note the good agreement between the growth exponent values obtained in our simulations and the predicted ones, equations \eqref{ew exponents} and \eqref{mh exponents} with $d=1$.

\begin{figure}[h]
\begin{center}

\includegraphics[width=8.1cm , height=5.4cm]{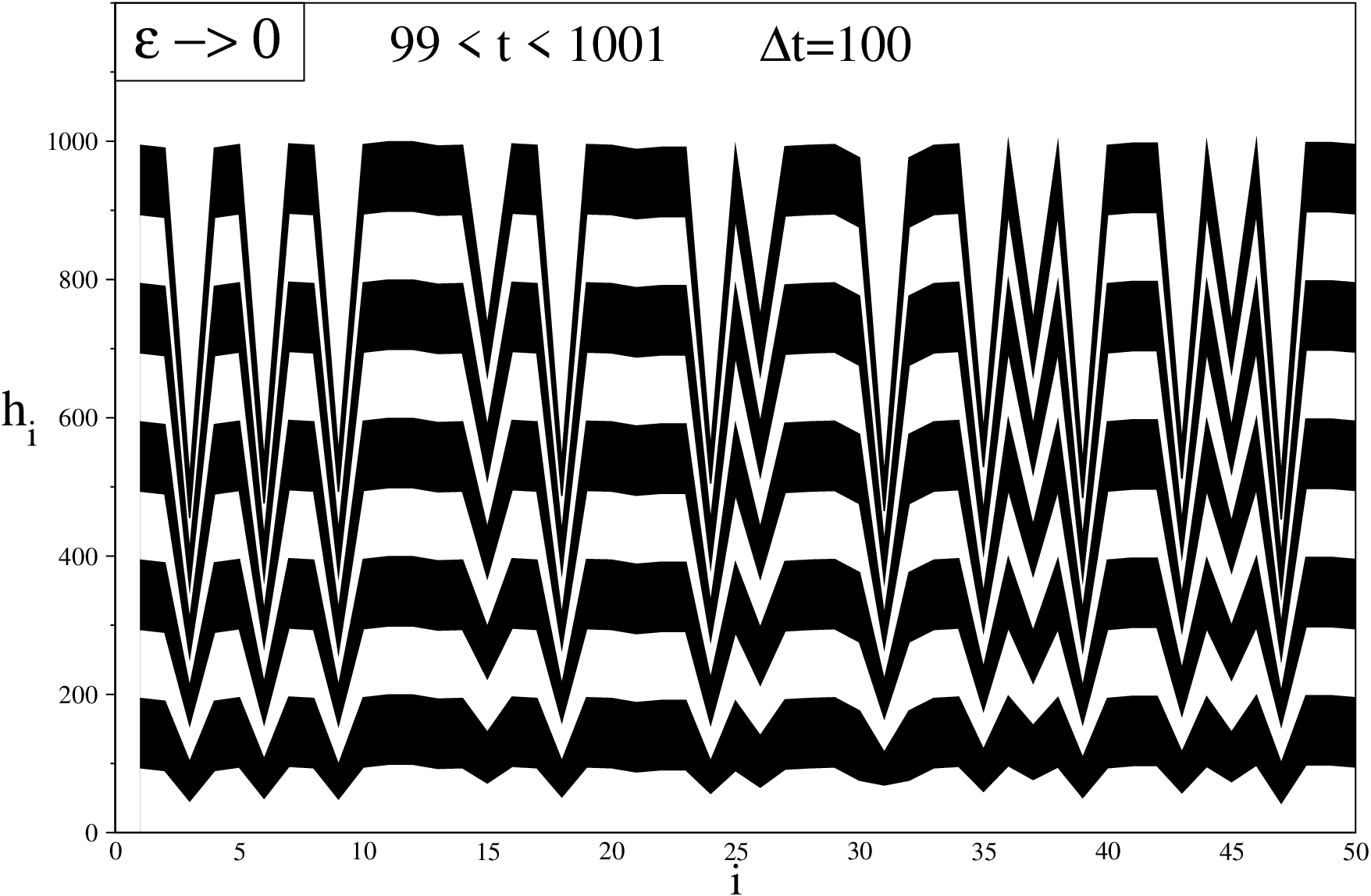}
\includegraphics[width=8.1cm , height=4.95cm]{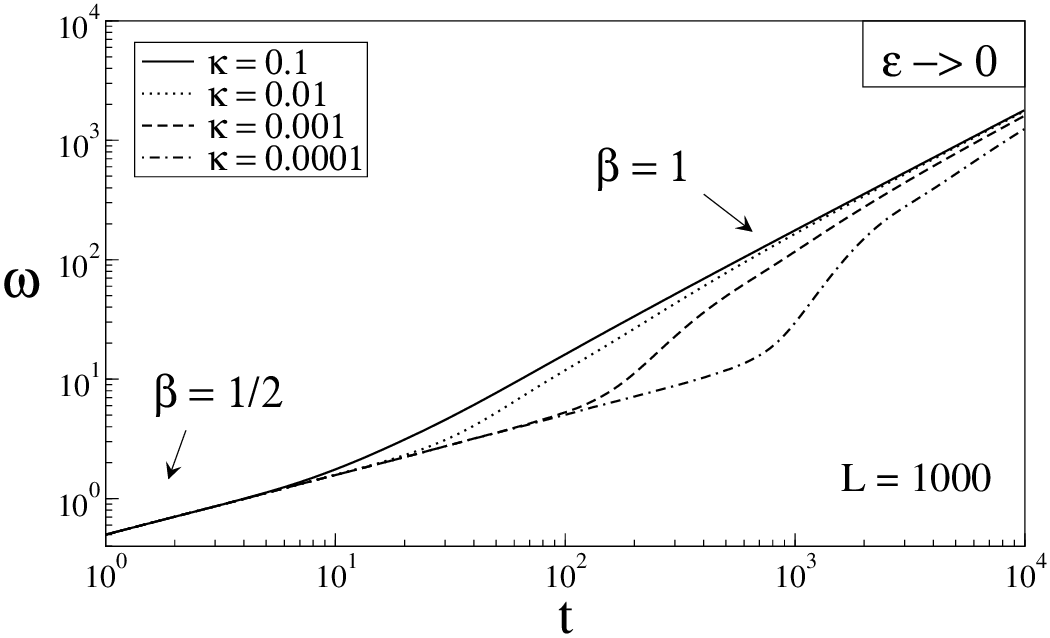}

\caption{\textit{\small In the top frame we show the evolution of the profiles by changing the color of the particles each $10^2$ time steps, for $10^2\leq t\leq10^3$, in the limiting case of $\varepsilon\rightarrow 0$. In the bottom we display the log-log plot of the time behavior of the roughness for $L = 10^3$ and several values of $\kappa$ between $10^{-1}$ and $10^{-4}$. Note that for smaller values of $\kappa$ the crossover $\beta=1/2\rightarrow\beta=1$ occurs at larger times.}}

\label{e=0}

\end{center}
\end{figure}

In the application of the method to the non-linear case of the KPZ equation, our simulations have shown that in the situation where the contribution from the Laplacian vanishes ($\varepsilon\rightarrow 0$) and the system evolves following a pure non-linear dynamics, one obtains persistent unevennesses growing between ``valleys'' and ``hills'', because the square of the gradient induces the development of such structures. We can see in the top frame of figure \ref{e=0} the evolution of a profile generated in the pure non-linear case, where the set of probabilities reaches a stable configuration such that $p_i=$ constant for all $i$: sites with larger heights have $p_i=1$ and sites in the valleys have $p_i=\rho$ (these valleys are symmetric, i.e. both sides have same slope). Thus, when that stage is attained by the system, the roughness, which is a measure of the width of the interface and can be thought as the difference between maximum and minimum heights, grows linearly with time and we get $\beta =1$. This behavior is shown in the bottom of figure \ref{e=0}. It is worthy to mention that this behavior is independent of the size of the system and occurs for all $\kappa$, the crossover from $\beta=1/2$ to $\beta=1$ happening for larger times as $\kappa$ decreases.

\begin{figure}[h]
\begin{center}

\includegraphics[width=8.1cm , height=5.4cm]{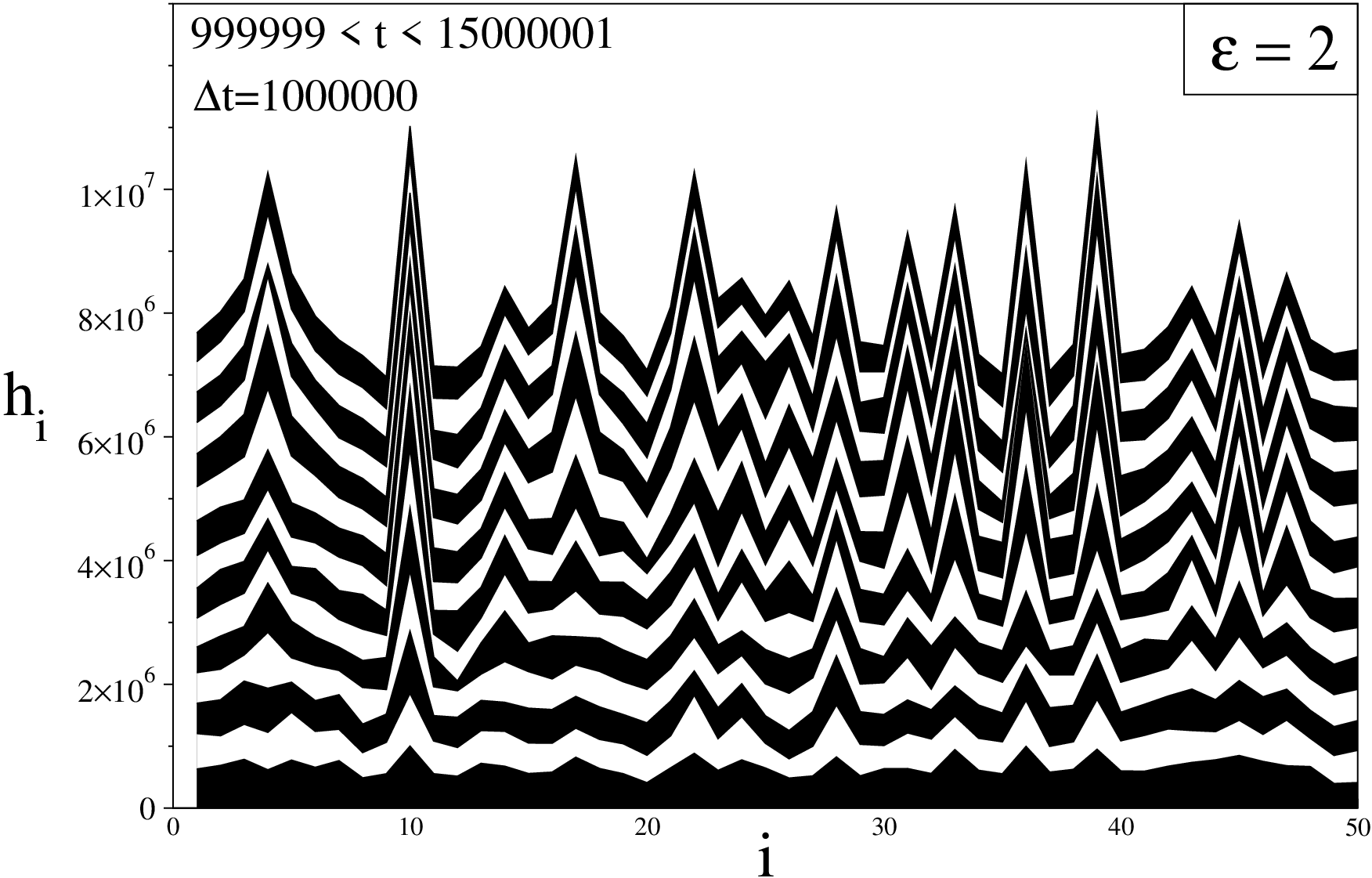}
\includegraphics[width=8.1cm , height=4.95cm]{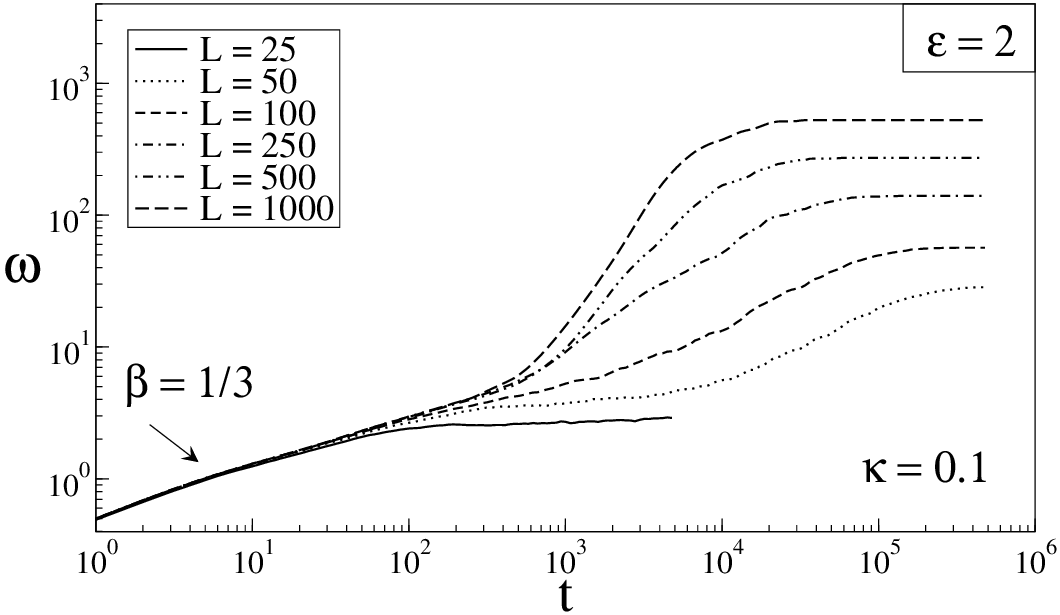}
\vspace{0.3cm}
\caption{\textit{\small In the top frame we show the interface evolution for the case of $\varepsilon =2$, for $10^6\leq t\leq 1.5\rm{x}10^7$, where we change the color of the particles each $10^6$ steps. In this case the system is very close to a crystallized pattern. In the bottom we have $\kappa =10^{-1}$ and several values of $L$ in the range $25 \leq L\leq 10^3$: larger systems reach crystallized patterns while smaller ones can saturate before that happens.}}

\label{e=2}

\end{center}
\end{figure}

In the top of figure \ref{e=2}, we show the profile generated when one sets $\varepsilon$ with small values ($\varepsilon =2$), which corresponds to a small, but non-vanishing, contribution from the Laplacian term. In this case the method produces crystallized patterns with asymmetric hills and valleys, in which all sites $i$ have $p_i(t)=1$ for all $t$ (after a certain transient time) and thus the roughness no longer changes. In the bottom of figure \ref{e=2} we show the behavior of the roughness for several system sizes and, as one can see, for larger systems the roughness grows initially with $\beta\approx 1/3$ before the frozen configuration is reached, while smaller systems can saturate before that. We believe that these structures are attractors among the possible configurations of heights in the profile. So, for small $\varepsilon$, the system always reaches such absorbing configurations. In our simulations the size of the system is restricted to $L\leq10^5$ so that we must make $\varepsilon > 4$ in order to have the standard behavior of the roughness and to avoid such anomalous configurations.

\begin{figure}[h]
\begin{center}

\includegraphics[width=8.89cm , height=8cm]{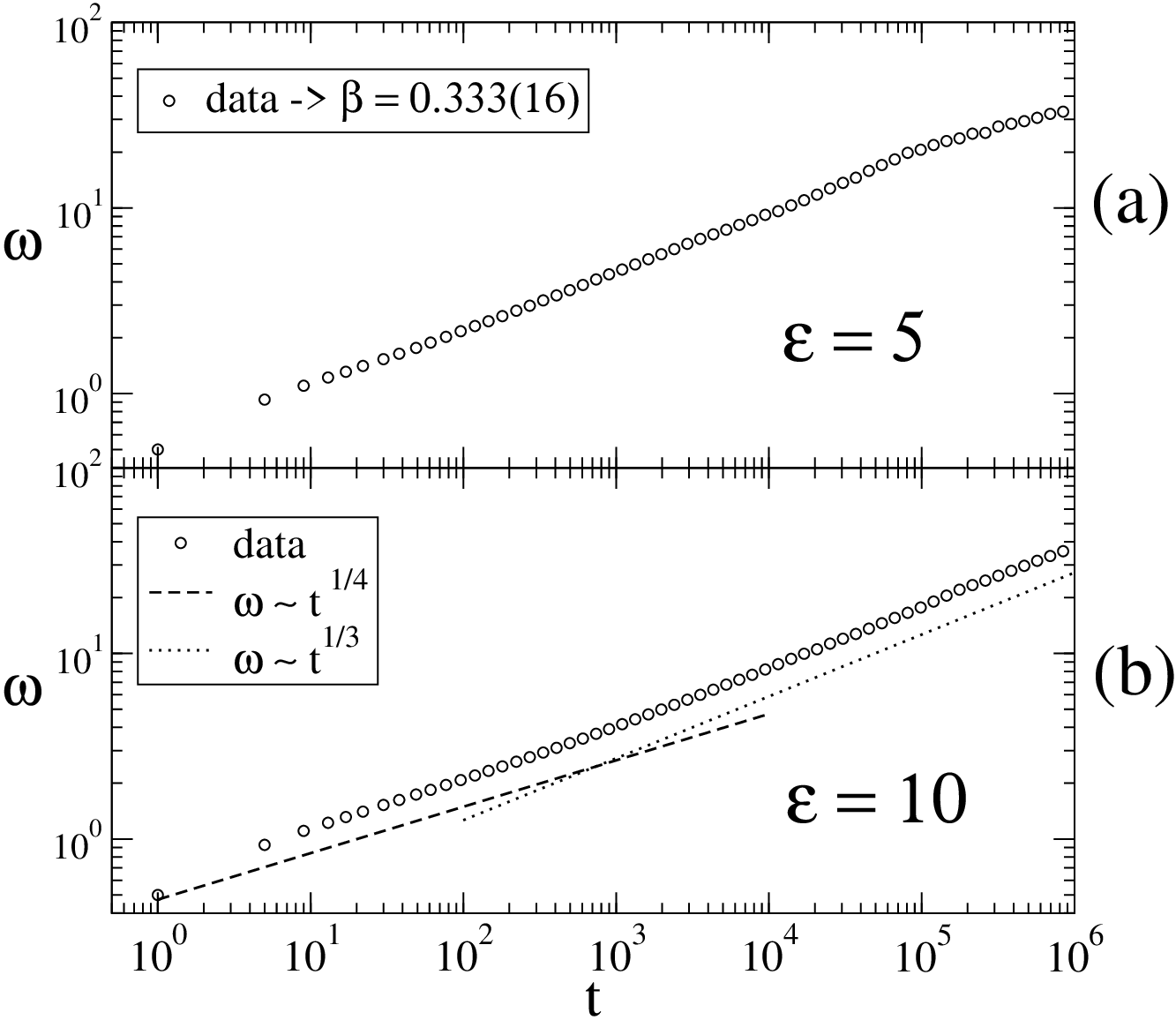}

\caption{\textit{\small Log-log plots of the roughness as function of time, in the application to the KPZ equation, where $\beta=1/3$ is expected. In the top, we have $\varepsilon=5$ for a system of size $L=10^4$. In the bottom, $\varepsilon=10$ and $L=10^5$, where a crossover from $\beta=1/4$ to $\beta=1/3$ is observed. We have drawn the functions $\omega\sim t^{1/4}$ (dashed line) and $\omega\sim t^{1/3}$ (dotted line) for comparison.}}

\label{rough kpz}

\end{center}
\end{figure}

We show in figure \ref{rough kpz}-a the results obtained for the roughness, with $\varepsilon=5$. Again, as one can see, a good agreement with prediction was obtained for the growth exponent $\beta$. By making $\varepsilon$ larger, a crossover from $\beta=1/4$ to $\beta=1/3$ regimes is observed, as one can see in figure \ref{rough kpz}-b, where we have $\varepsilon=10$ for a system of size $L=10^5$. This crossover occurs because in the beginning of the growth process, when the roughness is not large enough, the quadratic term of the KPZ equation is much smaller than the Laplacian, which dominates and provides $\beta\approx 1/4$. As the roughness increases, the non-linear term becomes dominant and we get $\beta\approx 1/3$. Of course, if we let $\varepsilon \rightarrow \infty$, this crossover does not occur anymore and we recover the results obtained for the EW class. This crossover has been previously obtained by da Silva and Moreira~\cite{tales} in a deposition model with restricted surface relaxation, where particles can relax only within a given distance $s$. If a minimum cannot be found in this range, the particle evaporates in a way similar to the RSOS growth model~\cite{kk}. An $s$-dependent crossover from $\beta=1/4$ to $\beta=1/3$ was obtained by the authors.

\begin{figure}
\begin{center}

\includegraphics[width=8cm , height=9.09cm]{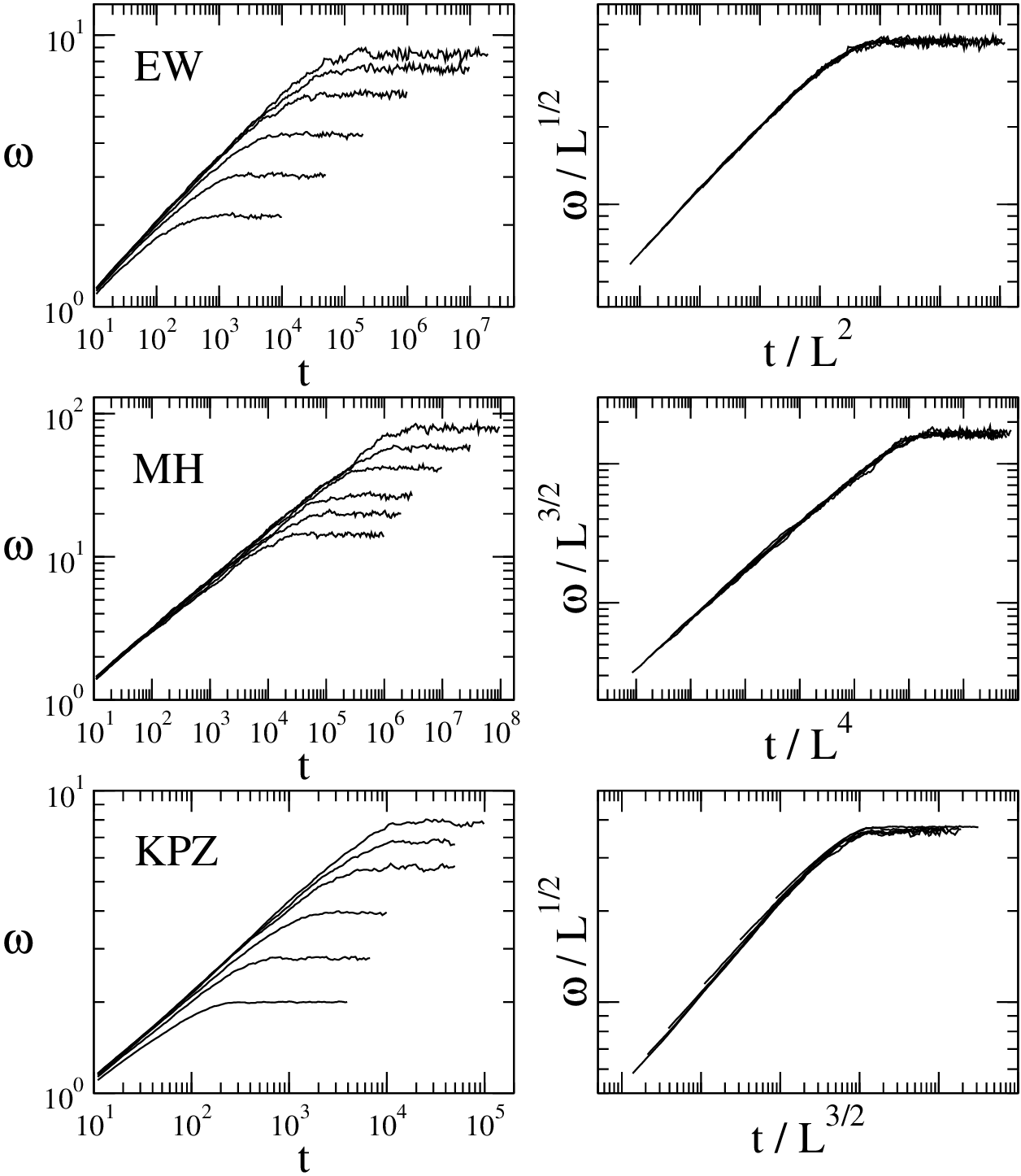}

\caption{\textit{\small Log-log plots of the roughness for various system sizes $L$, in the three applications of the method (left column). As we apply the Family-Vicsek scaling law, using the expected exponents for each UC, good collapses are obtained (right column).}}

\label{alpha & z}

\end{center}
\end{figure}

To determine the roughness exponent $\alpha$ and the dynamic exponent $z$, in order to have a complete characterization of each UC, we varied the size $L$ of the lattice, but still holding $\kappa=0.1$ fixed. We made $L=25$, $50$, $100$, $200$, $300$ and $400$ for the application to the EW and KPZ equations. In the MH class, for which $z=4$, we had to restrict our simulations to $L=20$, $25$, $30$, $40$, $50$ and $60$, due to the large saturation times for larger systems. The results are presented in figure \ref{alpha & z} where, in the left column, we show the roughness behavior for the several sizes $L$ considered. In the right column we apply the Family-Vicsek scaling law \eqref{scaling law}, with the corresponding expected value for $\alpha$ and $z$ for each UC, in order to obtain the collapse of the various curves into a single one. The good collapses obtained, together with the results for $\beta$, corroborate that our method indeed reproduces correctly each one of the three classes.

\section{Crossover from random to correlated growth} \label{crossover kappa}

As we vary the parameter $\kappa$, making it smaller, we identify a crossover from RD regime ($\beta=1/2$) to the corresponding correlated process. It was found that this crossover does not depend on the system size $L$.

Defining the crossover time as $t_c$, we see that $t_c$, $\omega_{sat}$ and $t_{\rm x}$ are all functions of the parameter $\kappa$, in a power law fashion:

\begin{equation} \label{lpt kappa}
\begin{array}{c}
t_c \sim \kappa^{-z^{\prime}_{\kappa}} ~,\\ \\
t_{\rm x} \sim \kappa^{-z_{\kappa}} ~,\\ \\
\omega_{sat} \sim \kappa^{-\alpha_{\kappa}} ~.
\end{array}
\end{equation}

\begin{figure}[h]
\begin{center}

\includegraphics[width=8cm , height=3.25cm]{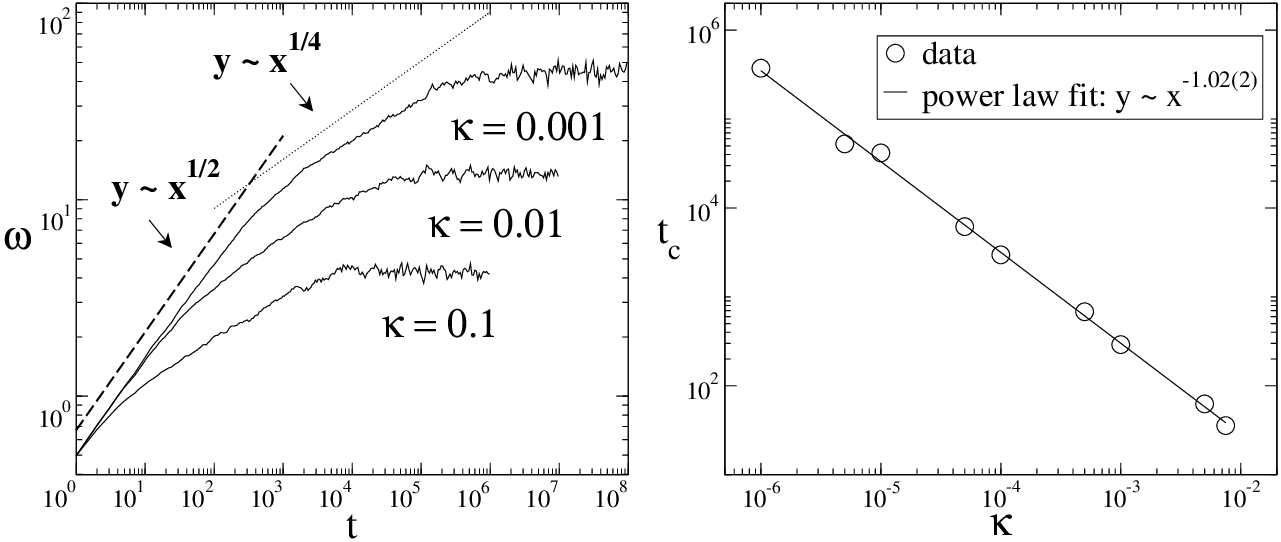}

\caption{\textit{\small In the left panel, the temporal behavior of the roughness for $L=250$ and $\kappa=10^{-1}$, $10^{-2}$ and $10^{-3}$, averaged over $40$ samples, in the application to the EW class. The crossover between $\beta = 1/2$ (RD) and $\beta = 1/4$ (EW) can be seen by comparing the curves with the dashed and dotted lines. In the right, the crossover time $t_c$ plotted against $\kappa$, exhibiting a power law with exponent $z^{\prime}_{\kappa}=1.02(2)$.}}

\label{tc}

\end{center}
\end{figure}

In the left of figure \ref{tc} we show the behavior of the roughness for different values of $\kappa$, in the EW application of the method: for a system of size $L=250$ we made $\kappa=10^{-1}$, $10^{-2}$ and $10^{-3}$, and averaged over $40$ independent samples. In the right of figure \ref{tc}, we show the curve $t_c \times \kappa$, where the power law fit provided $z^{\prime}_{\kappa}=1.02(2)$.

\begin{figure}[h]
\begin{center}

\includegraphics[width=8cm , height=3.25cm]{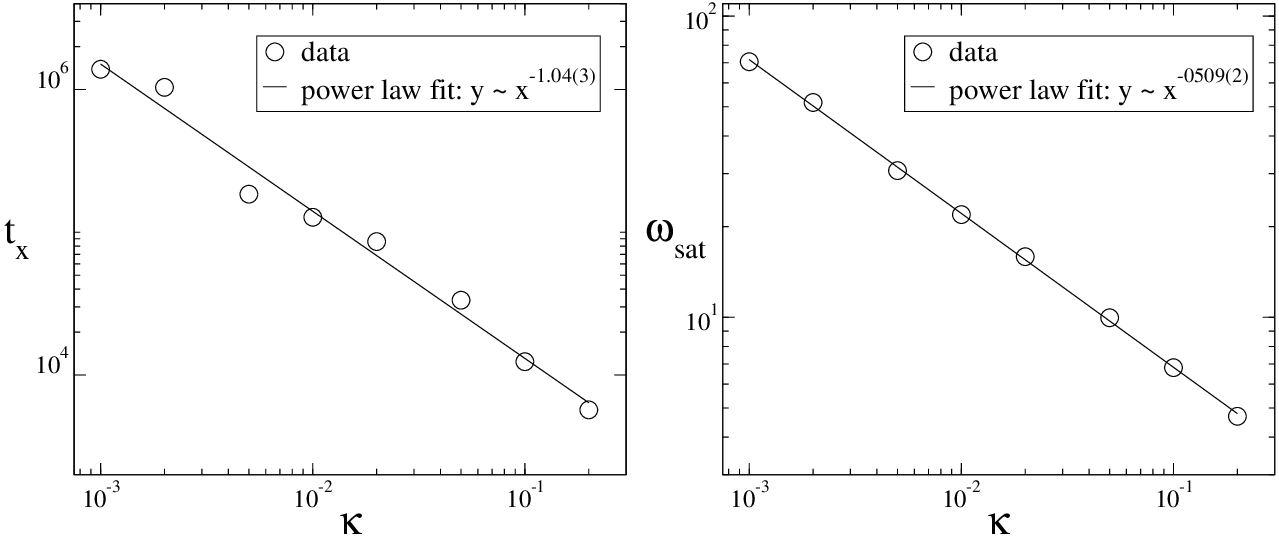}
\vspace{0.05cm}
\caption{\textit{\small Saturation time (left) and saturation roughness (right) as functions of the parameter $\kappa$, for a system of size $L=250$ and averaged over $40$ samples. From the power law fits we obtained $z_{\kappa}=1.04(3)$ and $\alpha_{\kappa}=0.509(2)$.}}

\label{tx wsat}

\end{center}
\end{figure}

In figure \ref{tx wsat}, we show the curves of saturation time and saturation roughness against $\kappa$, in the EW application. Once again, the system size is $L=250$ and the data is the average of $40$ independent samples. As one can see, we found $z_{\kappa}=1.04(3)$ and $\alpha_{\kappa}=0.509(2)$.

For the other two classes, MH and KPZ, we have found very close values for the $\kappa$-exponents as those obtained for the EW class. We conclude thus, that the crossover from random to correlated regime does not depend on the mechanism that generates correlations in the system. It is worthy to mention that for the KPZ class, when this crossover from random to correlated growth occurs, it is the Laplacian that dominates and the crossover should always be from random to linear correlated growth.

The fact that we have found $z_{\kappa}\cong z^{\prime}_{\kappa}\cong 1$ shows that $\kappa^{-1}$ plays the role of a characteristic time factor in the evolution of the system. Furthermore, $z^{\prime}_{\kappa}$ and $z_{\kappa}$ must have the same value because otherwise, making $\kappa$ small enough, we would have either the uncorrelated regime taking place over the correlated one (for the case $z^{\prime}_{\kappa} > z_{\kappa}$), or the correlated behavior stretching over and over (for $z^{\prime}_{\kappa} < z_{\kappa}$), which cannot happen unless the system size is increased. In other words, the quantity $t_{\rm x} - t_c$ is supposed to be a function only of the system size $L$.

The other result that we have obtained, $\alpha_{\kappa} \cong z^{\prime}_{\kappa}/2$, can be understood as follows. When $t=t_c$, the roughness value is, say, $\omega = \omega_c$ and, considering that so far the system is under the $\beta=1/2$ regime, we have $\omega_c = t_c^{1/2}$. It is also clear that $\omega_c \sim \omega_{sat}$, thus

\begin{equation}
\begin{array}{cc}
\omega_c \sim \omega_{sat} ~ \Longrightarrow ~ t_c^{\frac{1}{2}} \sim \omega_{sat} ~ \Longrightarrow ~  \kappa^{\frac{z^{\prime}_{\kappa}}{2}} \sim \kappa^{\alpha_{\kappa}} \\ \\
\Longrightarrow ~~ \alpha_{\kappa} = \dfrac{z^{\prime}_{\kappa}}{2}~.
\end{array}
\end{equation}


\section{Conclusions and perspectives} \label{conclusions}

We have introduced a new method, based on CA dynamics, to study stochastic differential equations associated to discrete deposition models. The method provides a new tool for obtaining the roughening exponents, which depends only on the discretization of the deterministic part of the equation, without having to actually solve it.

We applied this method to study two linear equations (EW and MH equations) and a non-linear one (KPZ equation), in $d=1$. The values obtained for the roughening exponents are in good agreement with prediction, showing that the method indeed reproduces each one of the three classes considered. In particular, for the non-linear case studied, a crossover from EW to KPZ class was obtained, for suitable values of parameter $\varepsilon$, which controls the non-linearity strength.

In addition, a crossover from RD to the considered correlated class was obtained when we varied the parameter $\kappa$. The crossover time, saturation time and saturation roughness were found to behave as power laws with $\kappa$, with numerical exponents $z^{\prime}_{\kappa}=1.02(2)$, $z_{\kappa}=1.04(3)$ and $\alpha_{\kappa}=0.509(2)$, respectively. These values have shown to be nearly the same, independently of the considered class.

In further works, we proceed by applying the method to growth equations in which other terms appear, such as $\nabla^2(\nabla h)^2$ and $\nabla\cdot(\nabla h)^3$, which are the corrections up to fourth order to the $\nabla^2 h$ term in the EW equation~\cite{lai das sarma}, as well as verifying the validity of the method to growth processes in two-dimensional lattices, where discretization schemes are not as trivial as in one dimension.
\\
\\
\\
The authors would like to thank F.D.A. Aarão Reis and D. Cavalcanti for helpful criticism on the manuscript. This work was supported by Brazilian agency CNPq.


\end{document}